\documentclass[aps,pre,superscriptaddress,twocolumn]{revtex4-2}

\usepackage{amsmath,amssymb,amsfonts}
\usepackage{graphicx}
\usepackage{bm}
\usepackage{hyperref}
\usepackage{xcolor}
\usepackage{grffile}
\usepackage{algpseudocode}  % Provides the algorithmic environment (from algorithmicx)
\usepackage{float} % Provides the [H] float specifier for better control
\usepackage{bm} % For better boldface in math mode

\newcommand{\cond}{\text{cond}}
\newcommand{\cross}{\text{cross}}

\newcommand{\algorithmpseudocode}[1]{%
  \renewcommand{\figurename}{Algorithm}%
  \caption{#1}%
  \addcontentsline{lof}{figure}{Algorithm: #1}%
  \renewcommand{\figurename}{Figure}% reset
}

\begin{document}

\title{The Longest Increasing Subsequence Problem revisited}

\author{Silvio Franz}
\email{silvio.franz@unisalento.it}
\affiliation{Dipartimento di Matematica e Fisica ``Ennio De Giorgi'', Universit\`a del Salento and INFN sezione di Lecce, via per Arnesano, 73100 Lecce, Italy}
\affiliation{Universit\'e Paris-Saclay, CNRS, LPTMS, 91405 Orsay, France}

\author{Roberto Mulet}
\email{roberto.mulet@gmail.com}
\affiliation{Group of Complex Systems and Statistical Physics and Department of Theoretical Physics, Faculty of Physics, University of Havana, CP 10400, La Habana, Cuba}

\date{\today}

\begin{abstract}
  The Longest Increasing Subsequence (LIS) problem—a classic combinatorial challenge with deep connections to statistical mechanics—exhibits a rich thermodynamic landscape. Introducing a temperature we identify two distinct energy scales: a Schottky-like crossover at $T_{\cross}$  and a condensation transition at $T_{\cond}$, below which the number of maximum-length configurations becomes sub-exponential in system size. We also show that despite the existence of polynomial-time dynamic programming algorithms for the ground state, local Monte Carlo dynamics, after sudden quenches at low temperatures, become trapped in metastable state configuration displaying characteristic glassy signatures: two-step relaxation, persistent dynamical overlaps, and aging. On the other hand, logarithmic annealing tracks equilibrium down to the ground state. These results establish that thermodynamic sparsity—not energetic barriers—can render local search dynamically intractable, positioning the LIS problem as a bridge between exactly solvable optimization and glassy spin-glass phenomenology.
\end{abstract}

\maketitle

\section{Introduction}

Given a sequence of distinct numbers, the Longest Increasing Subsequence Problem (LISP) seeks to determine the length of the longest subsequence that is strictly increasing. First suggested by Ulam in 1961~\cite{Ulam1961}, the LIS problem has since revealed connections at the intersection of mathematics~\cite{Bonomi2016}, computer science~\cite{Zhang2003}, and statistical physics~\cite{Prahofer2000}.

The asymptotic behavior of longest increasing subsequence lengths in random permutations is governed by the Baik--Deift--Johansson theorem~\cite{Baik1999}, which establishes a connection to random matrix theory: when properly centered and scaled as $\ell_{\max} = 2\sqrt{N} + N^{1/6}\chi$, the fluctuations converge to the Tracy--Widom distribution for the Gaussian Unitary Ensemble~\cite{Tracy1994}. This universality extends further---Pr{\'a}hofer and Spohn~\cite{Prahofer2000} demonstrated that LIS length fluctuations belong to the Kardar--Parisi--Zhang (KPZ).  The relevance of this result can not be underestimated, it connects the LISP to a  broad family of stochastic processes exhibiting identical scaling exponents and fluctuation statistics, including models of interface growth, stochastic surface dynamics, and one-dimensional asymmetric transport such as the asymmetric simple exclusion process (ASEP)~\cite{Kardar1986,Corwin2016}

Recent work has uncovered an even richer solution-space structure. Krabbe \textit{et al.}~\cite{Krabbe2020} showed that the entropy of longest increasing subsequences follows a Gaussian distribution with significant tail deviations. Surprisingly, in Ref.~\cite{Krabbe2023}, extending the problem to a thermodynamic framework, the authors interpret their results as signatures of replica symmetry breaking (RSB), a hallmark of complex energy landscapes typically associated with spin glasses and disordered systems~\cite{MPV}.

This apparent contradiction between KPZ universality and the possibility of replica symmetry breaking motivates our investigation into the thermodynamic properties of the LIS problem and, simultaneously, the dynamical behavior of local search algorithms designed to solve it. While directed polymers in random media (DPRM) in $1+1$ dimensions, a paradigmatic KPZ system, has no finite-T phase transitions~\cite{Derrida1990}, the possibility of an RSB transition suggests that correlations intrinsic to subsequence construction and the discrete combinatorial nature of LIS may yield fundamentally different behavior.

We follow   Krabbe's \textit{et al.}~\cite{Krabbe2020} and consider the length of increasing subsequences as an energy function. We then introduce a temperature parameter to probe the system's behavior across different energy scales. Our results reveal a complex phase diagram featuring two distinct temperatures: (i) we confirm the presence of a crossover at $T_{\cross} \approx 0.38$, and (ii) remarkably, a condensation transition at $T_{\cond} = 0.10 \pm 0.02$ where the system condensates into a sub-exponential number of ground states. Furthermore, we study also the dynamical behavior of the model under local Metropolis rules satisfying detailed balance and show that this temperature signal the appearance of glassy dynamics characteristic of complex systems.

%Since dynamic programming algorithms can efficiently find the ground state, this dual nature positions the LIS problem as a bridge between computationally tractable systems and those with complex energy landscapes.

\section{The Model}

Given a sequence of independent and identically distributed random numbers $\{s_i\}$ for $i=1,\dots,N$, Ulam's problem seeks the length of the longest increasing subsequence. An elegant formulation views the problem as a directed polymer on a random graph that can be solved in polynomial time by dynamic programming. Define $L(i)$ as the length of the longest increasing subsequence ending at position $i$, which satisfies the recurrence:
\begin{equation}
L(i) = 1 + \max_{\substack{1 \leq j < i \\ H_{ij}=1}} L(j),
\end{equation}
with $L(1)=1$, and where $H_{i,j} = \Theta(s_i - s_j)$ indicates whether $s_i > s_j$. The overall LIS length is then $L = \max_i L(i)$.

This formulation defines the set of Ulam random graphs: vertices at positions $(i, s_i)$ are connected by directed edges when $i < j$ and $s_i < s_j$, with $H_{i,j}$ as its adjacency matrix. Despite the graph's long-range connections, optimal paths use only local edges in index space, yielding a polymer length that scales as $\sim 2\sqrt{N}$~\cite{Borjes2019}.

Building on this graph formulation, we can count all increasing subsequences of length $\ell$ via dynamic programming~\cite{Krabbe2023}. Let $N_i^{\ell}$ be the number of such sequences ending at position $i$. These satisfy the recursion:
\begin{align}
N_i^{\ell} &= \sum_{j<i} H_{ij} N_j^{\ell-1}, \\
N_i^{1} &= 1.
\end{align}
or, in vector form, $\mathbf{N}^{\ell} = H^{\ell-1} \mathbf{N}^{1}$. Notice that $(H^{\ell-1})_{ij}$ counts directed paths of length $\ell-1$ from $j$ to $i$ in the Ulam graph. Consequently, the LIS length equals the largest $\ell$ for which $H^{\ell-1}$ is non-zero.

Conveniently, from the multiplicities \( N_i^\ell \), we can build a thermodynamic picture of the problem in the microcanonical formalism. Define an entropy \( S(\ell) \) as:
\begin{equation}
N(\ell) = e^{\sqrt{N} S(\ell/\sqrt{N},N)} = \sum_{i=1}^{N-\ell} N_i^\ell.
\end{equation}
and associate an energy \( E(\ell) = -\ell \) to each subsequence. In this microcanonical setting, the temperature and the specific heat can be computed as:
\begin{eqnarray}
&&    \frac{1}{T}=\frac{\partial S}{\partial E}\\
  && C_v=-\frac{1}{T^2\frac{\partial^2 S}{\partial E^2}}.
  \label{eq:cvmicro}
\end{eqnarray}
Notice that while the entropy is self-averaging in the thermodynamic limit, it fluctuates for finite $N$. Therefore to get the temperature as a function of energy with eq. (\ref{eq:cvmicro}), we measure the energy of each different sample relative to its ground state. This choice is crucial for identifying a condensation transition temperature.

Alternatively, in the canonical formalism, the partition function takes the form:
\begin{equation}
Z_N(\beta) = \sum_\ell N(\ell) e^{\beta \ell} = \sum_\ell e^{\sqrt{N} S(\ell/\sqrt{N},N)} e^{\beta \ell}.
\end{equation}
and the specific heat \( C_v \) can be computed as the variance of the energy \( E = -\ell \):
\begin{equation}
  c_v = \frac{\langle E^2 \rangle - \langle E \rangle^2}{\sqrt{N} T^2}
  \label{eq:CvvsT}
\end{equation}
where \( \langle \cdot \rangle \) denotes the average over multiple realizations of \( \sigma \). The reader must notice the $\sqrt{N}$ in the denominator needed to respect the scaling of the energy with the system size . We confront both, the microcanonical and the canonical, approaches in our numerical computations and found that they are equivalent.

Having presented the thermodynamic formalism, we now describe a local Monte Carlo algorithm in the canonical framework of fixed temperature. The dynamics satisfies detailed balance (see Appendix \ref{app:DBalance}) while preserving the increasing-subsequence constraint through elementary moves that add or remove elements. For a given temperature $T$, we propose new configurations using the move set in Algorithm~\ref{alg:mc-lis}

\begin{figure}[!htb]
  \begin{center}
      {\bf Local Algorithm: } Monte Carlo update satisfying detailed balance for the LISP at temperature $T$.
\vspace{0.5em}
    \end{center}
  \begin{algorithmic}[1]
\Require Random quenched sequence $A = (a_1, a_2, \dots, a_N)$, current increasing sub-se
quence $x_t$, temperature $T > 0$
\Ensure Updated sub-sequence $x_{t+1}$
\Statex
\State $x \gets x_t$ \Comment{Initialize working copy}
\For{$i = 1$ to $N$} \Comment{Perform $N$ elementary operations}
    \State Select $e \in A$ uniformly at random
    \If{$e \notin x$}
        \State Find all positions $j$ such that inserting $e$ at index $j$ in $x$ preserves increasing order
        \If{such positions exist}
            \State Insert $e$ at the first valid position in $x$ 
        \EndIf
    \ElsIf{$e \in x$ \textbf{and} $|x| > 1$}
        \State With probability $\exp(-1/T)$, remove $e$ from $x$
    \EndIf
\EndFor
\State $x_{t+1} \gets x$
\end{algorithmic}
  \algorithmpseudocode{Monte Carlo update satisfying detailed balance for the LISP at temperature $T$}
\label{alg:mc-lis}
\end{figure}

\section{Results}

\subsection{Thermodynamic Analysis}

\begin{figure}[!htbp]
\centering
\includegraphics[width=0.48\textwidth]{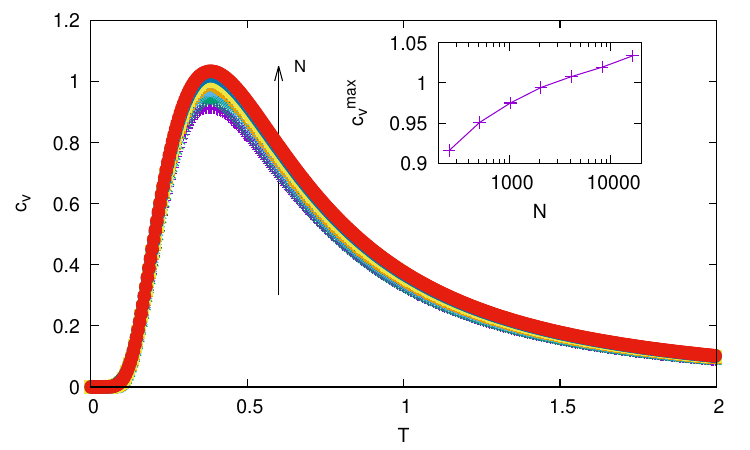}
\caption{Specific heat as a function of temperature for different system sizes $N=256,512,1024,2048,4096,8192,16384$. The peak at $T_{\cross} \sim 0.38$ indicates a finite-size crossover, while the behavior at lower temperatures reveals a freezing transition at $T_{\cond} = 0.10 \pm 0.02$.}
\label{fig:cv}
\end{figure}

\begin{figure}[!htb]
    \centering
    \includegraphics[width=1\linewidth]{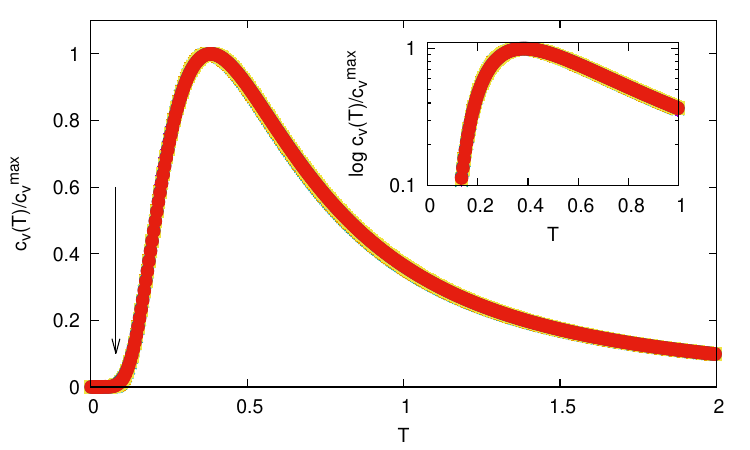}
    \caption{The specific heat $c_v(T,N)$ scaled by $c_v(N)^{max}$ for different system sizes $N=256,512,1024,2048,4096,8192,16384$. The arrow indicates the condensation transition. In the inset we reproduce the same plot on logarithmic scale to show the exponential decay at $T_{\cond} \approx 0.1$.}
    \label{fig:rescaled}
\end{figure}

To reveal the thermodynamic behavior we first look at the specific heat $c_v$. It is shown in Figure~\ref{fig:cv} (computed through Eq.~ \ref{eq:CvvsT} as a function of the temperature. It displays peak at $T_{\cross} \approx 0.38$ that grows logarithmically for growing values of $N$ (see the inset in Fig.~\ref{fig:cv}). The locus of the peak indeed coincides with the temperature below which Krabbe \textit{et al.}~\cite{Krabbe2023} report RSB-like effects. Moreover, in Figure \ref{fig:rescaled} we plot $c_v(T,N)$ divided by $c_v^{max}(N)$, and show that the curves for different $N$ perfectly collapse. This is a strong indication against a phase transition. We argue that the peak in the specific heat, instead, defines a Shottky crossover.

To make this Schottky picture precise, we can assume that the low-energy sector is described by $K(N)$ independent two-level systems with energy gap $\epsilon_0 = 1$. The partition function for each of these single systems is $Z_{\text{LIS}} = 1 + e^{-\beta\epsilon_0}$, and the intensive specific heat is:
\begin{equation}
%c_v(T) = \frac{K(N)}{\sqrt{N}} (\beta\epsilon_0)^2 \text{sech}^2\!\left(\frac{1}{2}\beta\epsilon_0\right).
c_v(T) = K(N) (\beta\epsilon_0)^2 \text{sech}^2\!\left(\frac{1}{2}\beta\epsilon_0\right).
\end{equation}

The peak occurs when $\beta\epsilon_0 \tanh(\frac{1}{2}\beta\epsilon_0) = 2$, whence the universal Schottky crossover temperature $T = 0.417$, close to our numerical simulations. Moreover, each position $k$ along the LIS is a candidate slot for an independent two-level system. Summing the exact combinatorial weights $w_k = 1/k+1/(\ell_{\max}-k+1)]$ yields $K(N) = 2H_{\ell_{\max}}/\ell_{\max} \sim \frac{1}{2}\ln N$ (see Appendix \ref{app:Kderivation}) that matches the numerical data (see the inset in ~\ref{fig:cv}).

Moreover, in  Figure \ref{fig:rescaled} $c_v$ goes to zero at finite Temperature as the system collapses into the ground states. The finite $N$ curves of $C_v$ as a function of $T$ display a last point at $T_0$, $C_v^*=C_v(T_0)$. We argue that for $N\to\infty$ there is a condensation transition at a finite temperature $T_{\cond}=0.10\pm 0.02$  where the system condenses in the ground state.

\begin{figure}[htbp]
\centering
\includegraphics[width=0.48\textwidth]{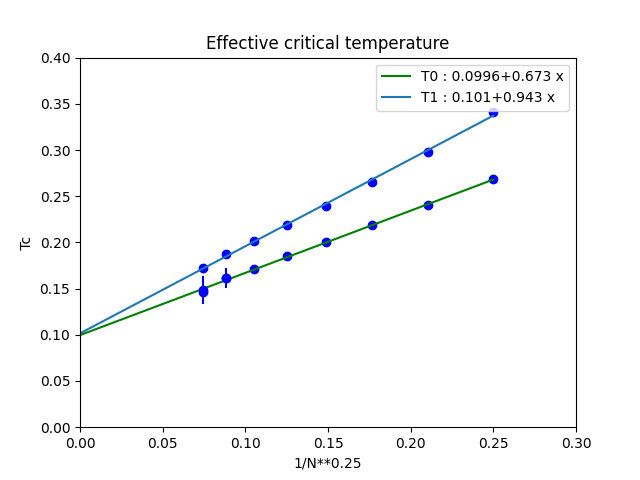}
\caption{The ground state and the first excited state temperature $T_0$ and $T_1$ respectively as a function of $1/N^{1/4}$. A linear fit indicates that both temperatures converge to $T_{\cond} = 0.10 \pm 0.02$ for $N \to \infty$, signaling a freezing transition in the thermodynamic limit.}
\label{fig:Tc}
\end{figure}

To see the condensation transition we study also the microcanonical temperature, $T(E)=1/(\partial S/\partial E)$   and define $T_0=T(E_0)$ from the right discrete derivative of $S$ in the ground state, $T_1=T(E_1)$ as the same thing for the first excited state. In Figure \ref{fig:Tc} we plot these quantities as a function of $N$.We see that plotted as a function of $N^{-1/4}$ both $T_0$ and $T_1$ are consistent with a linear dependence. Both temperatures appear to behave as $T_{0,1}=T_{\cond}+A_{0,1}/N^{1/4}$, and extrapolate to a common value $T_{\cond}=0.10\pm0.02$ for $N\to\infty$.

\begin{figure}[htbp]
\centering
\includegraphics[width=0.48\textwidth]{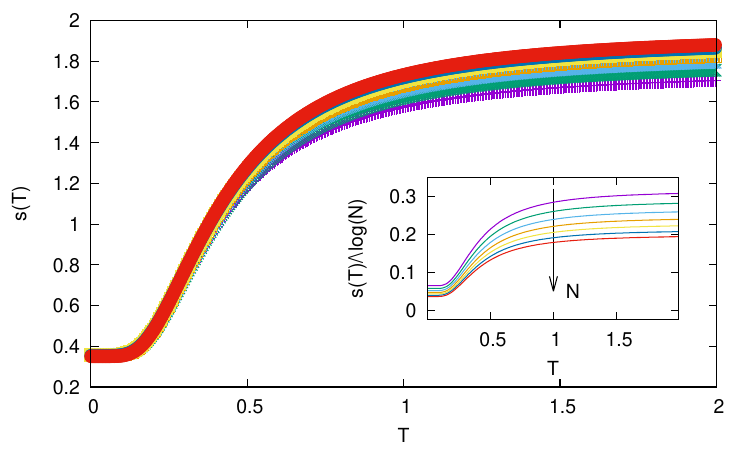}
\caption{Main panel: entropy density $s(T)=S(T)/\sqrt{N}$ versus temperature 
for system sizes $N=256,512,1024,2048,4096,8192,16384$. The collapse demonstrates that the configurational entropy scales as $S\sim\sqrt{N}$ in the condensed phase.  The low-temperature plateau $s_0\approx0.35$ reflects the residual entropy  of the ground-state manifold. Inset: the same data plotted as $s/\ln N$.  The systematic downward drift with $N$ rules out ordinary condensation with  positional entropy $S\sim\ln N$.}
\label{fig:entropy}
\end{figure}

The nature of this condensed phase is clarified by studying the scaling
of the configurational entropy.  Figure~\ref{fig:entropy} shows the
entropy density $s(T)=S(T)/\sqrt{N}$.  Below $T_{\rm cond}\approx 0.1$,
the curves for different $N$ converge to a common plateau $s(T)\to s_0\approx 0.35$, establishing that the number of ground-state configurations grows sub-exponentially as  $\Omega_{\rm gs}(N)\sim e^{s_0\sqrt{N}}$.

Above $T_{\rm cond}$, the curves do not collapse: $s(T)$ increases systematically with $N$, reflecting the exponential proliferation of sub-optimal configurations.  The inset in Figure~\ref{fig:entropy} makes this explicit by plotting $s/\ln N$.  In standard condensates (e.g.~BEC or zero-range processes) the low-temperature phase has positional entropy $S \sim\ln N$, which would produce $N$-independent plateaus in this representation.  Instead, the curves separate as $1/\ln N$, confirming that the condensed phase has far more degeneracy than ordinary condensation. %The presence of a finite residual entropy $s_0$ resolves the apparent entropy crisis: the system does not collapse onto a unique ground state ($S\to 0$) nor onto a logarithmically degenerate manifold ($S\sim\ln N$), but rather onto a sparse ground-state manifold whose cardinality scales as $e^{O(\sqrt{N})}$.

\begin{figure}[!htbp]
\centering
\includegraphics[width=0.48\textwidth]{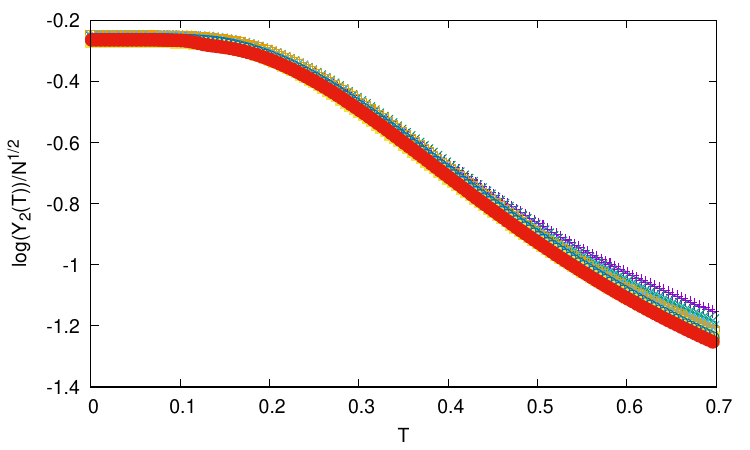}
\caption{Typical participation ratio $\log(Y_2)$ versus temperature $T$ for various system sizes, $N=256,512,1024,2048,4096,8192,16384$. The participation ratio measures the effective number of dominant configurations. Notice the collapse of the curves for $T<T_{\cond}$.}
\label{fig:y2}
\end{figure}

To further characterize this condensation, we compute the participation ratio:
\begin{equation}
Y_2 = \sum_{\alpha} p_\alpha^2,
\end{equation}
where $p_\alpha$ is the Gibbs weight of configuration $\alpha$.  For a
measure uniform over $\mathcal{N}_{\rm eff}$ states, $Y_2\approx 1/\mathcal{N}_{\rm eff}$ and for a delta-function measure, $Y_2=1$. In 
Fig.~\ref{fig:y2} we show that the curves $\log(Y_2)/\sqrt{N}$ for different $N$ collapse below $T_{\rm cond}\approx 0.1$.
This immediately rules out polynomial scaling $Y_2\sim N^{-\alpha}$ (which
would produce only an $O(1)$ vertical separation for the sizes shown).
Instead, the decay is consistent with a stretched exponential
$ Y_2(T)\sim\exp\!\bigl[-\Sigma_{\rm eff}(T)\sqrt{N}\,\bigr],$  where the effective complexity $\Sigma_{\rm eff}(T) = -\frac{\ln Y_2(T)}{\sqrt{N}}$ is a finite, $N$-independent quantity in the thermodynamic limit.

The plateau value $\Sigma_{\rm eff}\approx 0.25$ lies below the ground-state entropy density $s_0\approx 0.35$ obtained from Fig.~\ref{fig:entropy}. This inequality $\Sigma_{\rm eff}<s_0$ reflects a non-uniform Gibbs measure within the ground-state manifold: while $\Omega_{\rm gs}\sim e^{s_0\sqrt{N}}$ maximum-length subsequences exist, only $N_{\rm eff}\sim e^{\Sigma_{\rm eff}\sqrt{N}}$ of them carry significant weight. The gap $s_0-\Sigma_{\rm eff}\approx 0.10$ measures the entropic barriers that trap the dynamics inside a subset of the ground-state manifold, providing a quantitative link between the thermodynamic condensation and the glassy behavior discussed below. This sub-exponential condensation is intrinsic to the
combinatorial structure of the Ulam graph.  %Unlike disordered systems where condensation is diagnosed via replica-theoretic complexity functions, the LIS problem admits exact enumeration of paths via dynamic programming.

\subsection{Consequences for the Dynamics}

\begin{figure}[!htbp]
\centering
\includegraphics[width=0.48\textwidth]{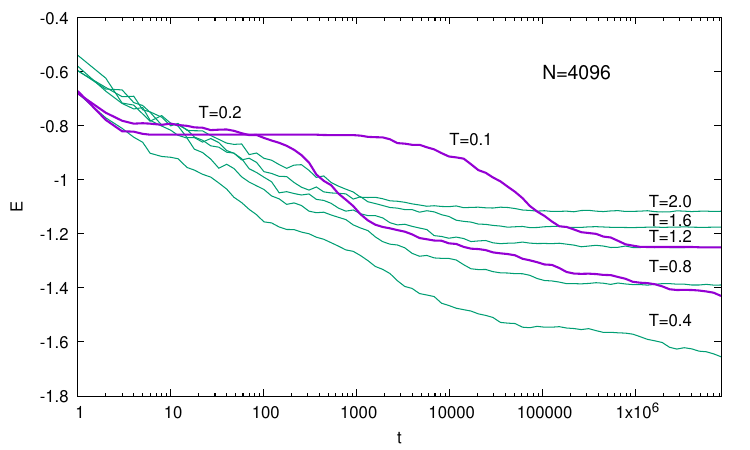}%fig_energy_time}
\caption{Energy versus time for Monte Carlo simulations at various temperatures, $N=4096$. The dramatic slowing down below $T_{\cross}$ is indicative of the appearance of a complex landscape.}
\label{fig:energy_time}
\end{figure}

This thermodynamic picture has  dynamical consequences for the local search {\bf Algorithm:}~\ref{alg:mc-lis}. Figure ~\ref{fig:energy_time} shows the time evolution of the energy of the system ($E= -\ell$) after a sudden quench from from infinite temperature to a target $T$.

At high temperature ($T > T_{\cross} \approx 0.4$) the energy relax exponentially to their equilibrium values. In this region of temperatures the ground-state manifold contains exponentially many LIS configurations, and the Gibbs factor $e^{\beta\epsilon_0}$ is almost one; all the moves are accepted and the system reaches equilibrium very fast. When $T_{\cond} < T < T_{\cross}$ the system first relaxes rapidly within the degenerate manifold, then the tunneling between distinct LIS configurations requires collective rearrangements and the system slows down showing a characteristic two step-relaxation(see Fig.~\ref{fig:energy_time}).  At $T=0.1$, the system fails to reach equilibrium within the observation window.

\begin{figure}[!htb]
    \centering
    \includegraphics[width=0.45\textwidth]{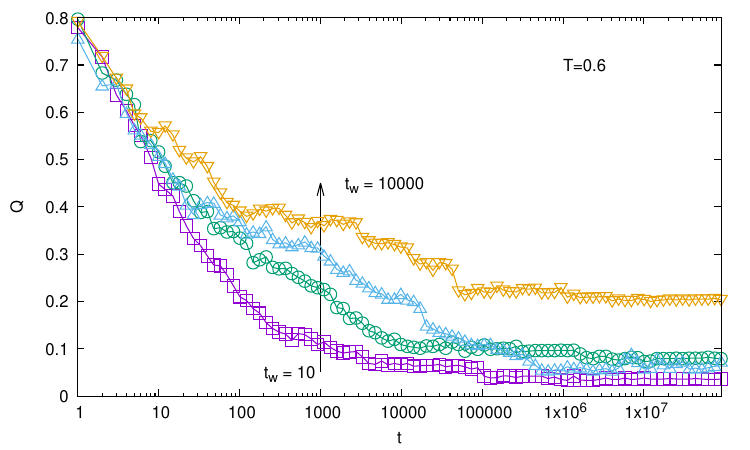}
    \includegraphics[width=0.45\textwidth]{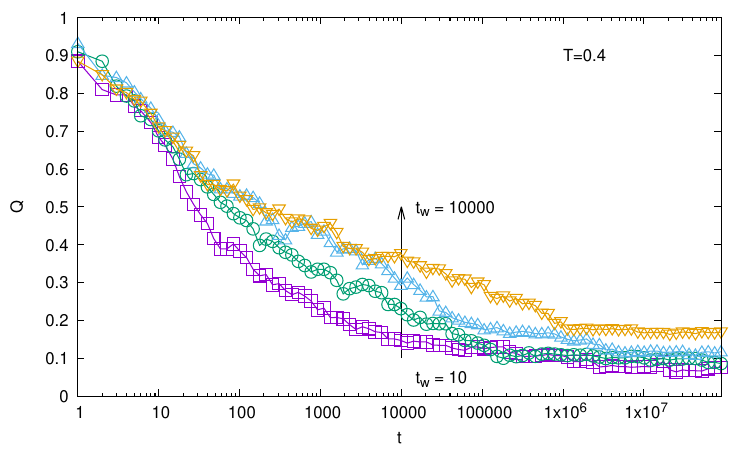}
    \includegraphics[width=0.45\textwidth]{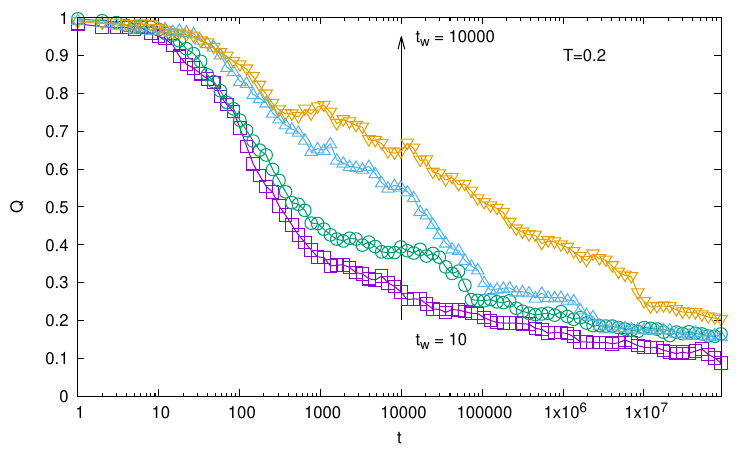}
    \includegraphics[width=0.45\textwidth]{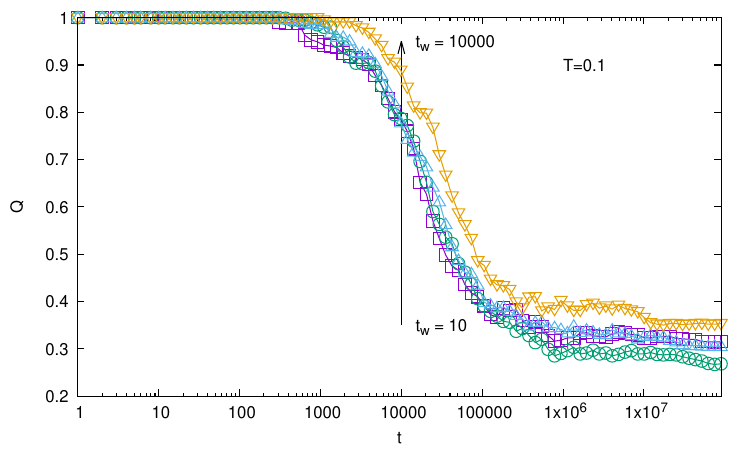}
    \caption{Overlap versus time for Monte Carlo simulations at temperatures \( T = 0.6, 0.4, 0.2, 0.1 \) and \( N = 1024 \), $t_w=10,100,1000$ and $10000$. The slowing down of overlap dynamics at $T \sim 0.4$ is also  indicative of the appearance of a complex landscape, while persistent overlaps at lower temperatures indicate the condensation transition.}
    \label{fig:Qvst}
\end{figure}

To probe the \emph{configurational} nature of this slowdown we study the dynamical overlap
\begin{equation}
Q(t)=\frac{\langle \mathbf{x}(t+t_w)\cdot\mathbf{x}(t_w)\rangle}{N},
\end{equation}
where  $x(t)$ is the increasing subsequence at time $t$, represented as a binary vector with $x_i=1$ if element $i$ is selected and $\mathbf{x}(t_w)$ is the subsequence at waiting time $t_w$ after the quench. The results for $N=1024$ and $t_w=10,100,1000,10\,000$ are shown in Fig.~\ref{fig:Qvst}.

%This picture is mimicked by the time evolution of the dynamical overlap $Q(t)$, the fraction of elements shared with the initial subsequence. Defining $x(t)$ as the increasing subsequence at time $t$, represented as a binary vector with $x_i=1$ if element $i$ is selected we define $Q(t)=\langle x(t)\cdot x(0)\rangle/N$, where the dot product counts the number of common elements (shared indices/values) between the current subsequence $x(t)$  and the initial subsequence $x(0)$.

At $T=0.6>T_{\rm cross}$ the energy equilibrates rapidly (cf.~Fig.~\ref{fig:energy_time}), yet the overlap displays a pronounced $t_w$ dependence. The curves start at $Q\approx 0.8$---fast local moves decorrelate the configuration within the first Monte Carlo sweep---and then decay continuously. For short waiting times ($t_w=10$) the overlap falls to $Q\approx 0.05$ by $t\sim 10^3$, indicating that the system quickly wanders away from its initial configuration. For long waiting times ($t_w=10\,000$) the curve plateaus at $Q\approx 0.2$ for the entire observation window. This persistent memory is not caused by energetic barriers, the specific heat is featureless here and the system is thermodynamically simple, but by the sheer entropic size of the solution space. During $t_w$ the system ages into similar paths of mutually accessible configurations, and subsequent dynamics mostly shuffles elements between these path without escaping from them. The $t_w$ dependence at $T=0.6$ is already a signature of \emph{entropic} trapping. A similar continuous decay is seen at $T=0.4\approx T_{\cross}$, though the overall relaxation is slower and the initial overlap is higher ($Q\approx 0.9$). The $t_w$ separation remains strong, confirming that even at the Schottky crossover the dominant obstacle for local search is the vastness of the degenerate manifold, not merely the two-level excitations discussed above. At $T=0.2$ ($T_{\cond}<T<T_{\cross}$) the dynamics exhibit a clear separation of timescales. For $t\lesssim 10$ all curves collapse and stay together independently of $t_w$.  For $t\gtrsim 10^2, 10^3$ the curves separate dramatically, with the overlap decaying much more slowly for larger $t_w$. This system experiments inter-basin exploration performing collective rearrangements to find a structurally different subsequence, and longer waiting times trap it deeper within the same set of metastable paths.

The behavior changes qualitatively at $T=0.1\approx T_{\cond}$. Here the overlap remains locked at $Q\approx 1$ for $t\lesssim 10^3$, then undergoes a sharp, almost synchronized drop to a high persistent plateau ($Q\approx 0.3$--$0.4$). Crucially, the $t_w$ dependence is now much weaker than at $T=0.2$: the four curves are far closer together, with only a modest delay of the drop and a slight increase of the final plateau for larger $t_w$. The weakening of the waiting-time effect indicates that the system has condensed into a sparse, effectively frozen subset of the ground-state manifold. Unlike the high-temperature entropic trapping, where longer $t_w$ progressively deepens the memory, the condensed phase is dynamically arrested: the Gibbs measure concentrates on only $N_{\rm eff}\sim e^{\Sigma_{\rm eff}\sqrt{N}}$ dominant configurations (cf.~Fig.~\ref{fig:y2}). At this temperature local Monte Carlo moves cannot drive the system to a structurally different dominant state regardless of how long it waits.

This progression, from entropic slow mixing at high temperatures, through strong aging in the complex landscape at $T=0.2$, to condensation-induced arrest at $T=0.1$---establishes that the dynamical hardness of the LISP for local search has two distinct origins. Above $T_{\cross}$ the obstacle is the exponential size of the solution space explored by local moves; below $T_{\cond}$ it is the thermodynamic sparsity of the Gibbs measure.

\begin{figure}[!htbp]
\centering
\includegraphics[width=0.48\textwidth]{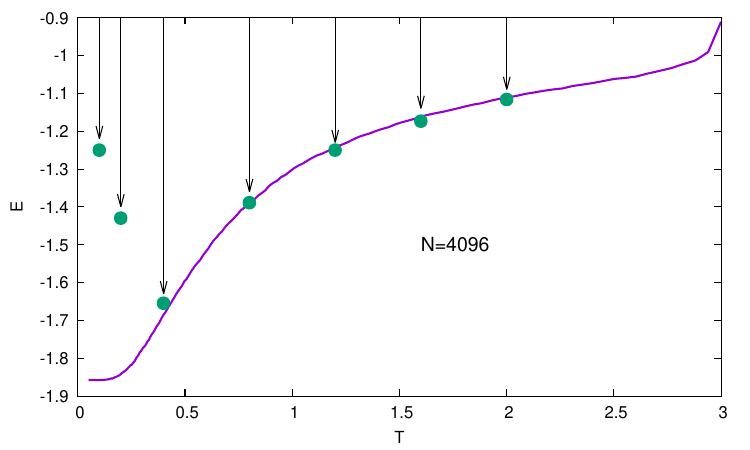}%fig_annealing}
\caption{Energy versus temperature for $N=4096$. The line represents results from simulated annealing, while points represent the final energy obtain after a quench to fixed temperatures. The arrows are guides to the eyes. Simulated annealing finds the ground state, while quenches below $T_{\cond}$ get trapped in metastable states.}
\label{fig:annealing}
\end{figure}

On the other hand, there is an evident contrast between the performance of the algorithm after a quench from high temperatures and simulated annealing. As shown in Fig.~\ref{fig:annealing}, the annealing trajectory tracks the equilibrium curve down to the ground state, whereas quenches at $T < T_{\cond}$ plateau at higher energies.

A logarithmic annealing schedule restores equilibrium dynamics and allows the algorithm to reach the ground state. This may look surprising since the system condensates at $T_{\cond} \approx 0.10$. The intuition is that simulated annealing succeeds because it builds subsequences monotonically while $T > T_{\cond}$ (where dynamics remains fast), reaching $\ell \approx \ell_{\max}$ before encountering the sparsity-induced slowdown. A sudden quench, by contrast, crosses $T_{\cond}$ at small $\ell$, freezing the dynamics before optimality is achieved. This supports the idea that the thermodynamic sparsity of the solution space---not energetic barriers---generates algorithmic hardness in this problem. The success of annealing is not merely a matter of patience; it is a qualitatively different strategy that exploits the temperature-dependent structure of the landscape. At high $T$, the entropy-rich regime allows rapid exploration; the algorithm must discover the correct energy basin before the condensation transition removes the configurational diversity needed for further search.

\section{Conclusions}

We have shown that the thermal version of the Longest Increasing Subsequence problem---a paradigmatic exactly solvable combinatorial optimization problem---exhibits a rich finite-temperature phenomenology governed by two distinct energy scales.

At $T_{\cross} \simeq 0.38$, the system exhibits a Schottky-like crossover driven by $O(\ln N)$ independent two-level systems associated with active gaps along the LIS backbone. At lower temperatures, $T_{\cond} \simeq 0.10 \pm 0.02$, the manifold of increasing subsequences undergoes a condensation transition onto a sub-exponential set of maximum-length configurations, signaled by the saturation of the entropy density $S/\sqrt{N}\to s_0\approx 0.35$. These phenomena are lattice echoes---artifacts of the microscopic discreteness inherent to the Ulam graph formulation. The locked excitation gap $\epsilon_0 = 1$ and the combinatorial constraints on subsequence construction generate an effective rugged energy landscape. This would vanish entirely in the continuum limit ($\epsilon_0 \to 0$), where standard KPZ universality prevails. Thus, the discrete Ulam lattice interpolates between integrable KPZ statistics at high energies and glassy dynamics at low energies.

Our work exposes a striking algorithmic paradox. Despite the existence of polynomial-time dynamic programming algorithms for finding the ground state, local Monte Carlo dynamics exhibits characteristic signatures of glassiness: two-step relaxation, persistent dynamical overlaps, and logarithmic annealing protocols required to avoid metastable traps. The condensation transition at $T_{\cond}$ represents a fundamental limit where thermodynamic sparsity---not energetic barriers---generates algorithmic hardness. In short: computational complexity and physical glassiness are distinct concepts: a problem may be formally tractable yet dynamically intractable for local search algorithms. From this point of view, the LISP provides a canonical bridge between easy optimization problems and hard spin-glass systems.%, demonstrating that complex landscapes and sluggish dynamics can emerge purely from microscopic discreteness and combinatorial structure, independent of quenched disorder. This suggests that glassy phenomenology may be far more ubiquitous in combinatorial problems than previously recognized.

\appendix

\section{Detailed Balance}
\label{app:DBalance}

Let $x$ and $y$ be valid increasing subsequences with lengths $m$ and $m+1$ respectively, such that $y = x \cup \{e\}$ for some element $e \notin x$ that can be validly inserted.

The transition probability $P(x \to y)$ consists of:
\begin{itemize}
    \item Probability to select element $e$: $\frac{1}{N}$
    \item Probability of successful insertion: 1 (due to exhaustive search)
    \item Acceptance probability: 1
\end{itemize}
Thus: $P(x \to y) = \frac{1}{N}$

The reverse transition $P(y \to x)$ consists of:
\begin{itemize}
    \item Probability to select element $e$: $\frac{1}{N}$
    \item Removal probability: $\exp(-1/T)$
\end{itemize}
Thus: $P(y \to x) = \frac{1}{N} \exp(-1/T)$

The detailed balance condition requires:
\[
\pi(x) P(x \to y) = \pi(y) P(y \to x)
\]

Substituting the Boltzmann distribution $\pi(x) \propto \exp(|x|/T)$:
\begin{align*}
\exp\left(\frac{m}{T}\right) \cdot \frac{1}{N} &= \exp\left(\frac{m+1}{T}\right) \cdot \frac{1}{N} \cdot \exp(-1/T) \\
\exp\left(\frac{m}{T}\right) &= \exp\left(\frac{m+1}{T}\right) \cdot \exp(-1/T) \\
\exp\left(\frac{m}{T}\right) &= \exp\left(\frac{m}{T}\right) \cdot \exp(1/T) \cdot \exp(-1/T) \\
\exp\left(\frac{m}{T}\right) &= \exp\left(\frac{m}{T}\right)
\end{align*}

The identity holds, confirming that detailed balance is satisfied. The removal probability $\exp(-1/T)$ exactly compensates for the Boltzmann factor ratio $\exp(1/T)$ between states differing by one element.

This algorithm provides an efficient Markov chain Monte Carlo method for sampling from the distribution $\pi(x) \propto \exp(|x|/T)$ while maintaining the increasing subsequence constraint throughout the evolution.

\section{Microscopic Derivation of $K(N)$}
\label{app:Kderivation}

In the Schottky description of the low--temperature thermodynamics, the intensive specific heat is written as
\begin{equation}
c_v(T)=K(N)\,(\beta\epsilon_0)^2\operatorname{sech}^2\!\Bigl(\frac{\beta\epsilon_0}{2}\Bigr)\;,
\qquad \epsilon_0=1\; .
\label{eq:schottky}
\end{equation}
For a single isolated two--level system one would have $K(N)=1$ and the peak height would be a pure number.  Numeric show instead a peak that grows slowly with system size as $c_v^{\max}\sim\ln N$.  Hence the system must behave as a collection of $K(N)$ effective independent two--level systems with
\begin{equation}
K(N)\sim \ln N\; .
\label{eq:Ktarget}
\end{equation}

Below we derive this law microscopically by counting the effective binary degrees of freedom associated with the backbone of a longest increasing subsequence.

In the Schottky phenomenology the low-energy sector is described by $K(N)$ independent two-level systems. Each system corresponds to a \emph{slot} $k \in \{1, \dots, \ell_{\max}\}$ along the LIS backbone that hosts a \emph{binary choice}:
\begin{itemize}
    \item \textbf{State 0:} the slot is filled in a way compatible with a maximum-length subsequence (energy $E = 0$);
    \item \textbf{State 1:} the slot is blocked or deformed, reducing the achievable length by one (energy $E = +\epsilon_0$ with $\epsilon_0 = 1$).
\end{itemize}
A slot contributes to $K(N)$ only if this binary choice is \emph{sharp}---that is, if the slot is \emph{rigid}, meaning the element sitting there cannot be replaced by an unused alternative without destroying the LIS property.

In the Schottky phenomenology the low-energy density of states is approximated as
\begin{equation}
\Omega(E_{\rm gs}+m)\;\approx\;\Omega_{\rm gs}\binom{K(N)}{m}\;,
\qquad m\ll K(N)\; ,
\end{equation}
where $\Omega_{\rm gs}$ is the ground-state degeneracy.  Setting $m=1$ gives the exact microscopic identification
\begin{equation}
K(N)\;=\;\frac{\Omega(E_{\rm gs}+1)}{\Omega_{\rm gs}}
\;=\;\frac{N(\ell_{\max}-1)}{N(\ell_{\max})}\; .
\tag{B.1}
\end{equation}
Here $N(\ell)$ is the total number of increasing subsequences of length $\ell$.  We compute this ratio by counting deletions from the LIS backbone. It is important first to introduce the concept of {\it Rigid slots}, i.e.: when  the gap $(s_{k-1},s_{k+1})$ contains no unused values.  In this case the deletion is irreversible and produces a \emph{unique} length-$(\ell_{\max}-1)$ subsequence that cannot be obtained from any other LIS.

%Every length-$(\ell_{\max}-1)$ subsequence can be obtained by deleting one element from some LIS.  However, different deletions may produce the same subsequence.  The key distinction is between:
%\begin{itemize}
 %   \item \textbf{Rigid slots:} the gap $(s_{k-1},s_{k+1})$ contains no unused values.  The deletion is irreversible and produces a \emph{unique} length-$(\ell_{\max}-1)$ subsequence that cannot be obtained from any other LIS.
  %  \item \textbf{Non-rigid slots:} the gap contains many unused values.  The deletion can be repaired in $\sim\!\sqrt{N}$ ways, so the same subsequence is produced redundantly by many LIS.
%\end{itemize}
%For a non-rigid slot, the redundancy $r\sim\sqrt{N}$ suppresses its contribution to $K(N)$ by $1/r\sim 1/\sqrt{N}$.  The net contribution of all non-rigid slots is $O(1)$, subleading compared to rigid slots.

Then, for a given LIS $\sigma$, let $R(\sigma)$ be the number of rigid slots.  Each rigid slot produces exactly one distinct first-excited state.  Averaging over all LIS and using linearity of expectation:
\begin{equation}
\Bigl\langle R \Bigr\rangle
\;=\;
\Bigl\langle \sum_{k=1}^{\ell_{\max}} \mathbf{1}[\text{slot }k\text{ rigid}] \Bigr\rangle
\;=\;
\sum_{k=1}^{\ell_{\max}} \mathbb{P}[\text{slot }k\text{ is rigid}]
\;=\;
\sum_{k=1}^{\ell_{\max}} w_k
\tag{B.2}
\end{equation}

Because rigid deletions are unique to one LIS, the total number of distinct first-excited states is
\begin{equation}
\Omega(E_{\rm gs}+1) \;=\; \Omega_{\rm gs}\cdot\Bigl\langle R \Bigr\rangle
\;=\; \Omega_{\rm gs}\sum_{k=1}^{\ell_{\max}} w_k\; .
\tag{B.3}
\end{equation}

Dividing by $\Omega_{\rm gs}$ and using (B.1): $K(N)\;=\;\sum_{k=1}^{\ell_{\max}} w_k\; .$, i.e.: $K(N)$ is the expected number of rigid slots per LIS.%, obtained by summing the rigid-slot probabilities.

%\begin{equation}
%K(N)\;=\;\sum_{k=1}^{\ell_{\max}} w_k\; .
%\tag{B.4}
%\end{equation}
%Thus $K(N)$ is the expected number of rigid slots per LIS, obtained by summing the rigid-slot probabilities.

For slot $k$ to be rigid, the local ordering constraints must be tight. Because the quenched sequence is i.i.d., the left and right constraints are statistically independent.

Consider the $k$ ordered backbone values on the left side of slot $k$:
\begin{equation}
    s_1 < s_2 < \cdots < s_k\;.
\end{equation}
These are the \emph{order statistics} of $k$ i.i.d.\ continuous random variables drawn from the quenched sequence. Before we condition on their ordering, they were just $k$ exchangeable values.

Among $k$ exchangeable continuous random variables, each is equally likely to be the largest. Therefore the probability that $s_k$ (the last in the ordered list) was the \emph{original maximum} is exactly
\begin{equation}
    \mathbb{P}\bigl[s_k = \max\{s_1, \ldots, s_k\}\bigr] = \frac{1}{k}\;.
    \label{eq:leftprob}
\end{equation}

When $s_k$ is the maximum of the first $k$ backbone values, the left side is maximally constrained. Any unused value that could replace $s_k$ would need to be larger than $s_{k-1}$ (to preserve the increasing order) but smaller than $s_k$ (to fit in the slot). Since $s_k$ is already the largest among $\{s_1, \ldots, s_k\}$, such values are scarce. The larger $k$ is, the more ``competition'' $s_k$ faces from preceding values, and the less likely it is to be the maximum---hence the $1/k$ suppression.

By the identical argument on the right side of slot $k$, there are $\ell_{\max}-k+1$ ordered values 
\begin{equation}
    s_k < s_{k+1} < \cdots < s_{\ell_{\max}}\;.
\end{equation}
and

\begin{equation}
    \mathbb{P}\bigl[s_k = \min\{s_k, \ldots, s_{\ell_{\max}}\}\bigr] = \frac{1}{\ell_{\max}-k+1}\;.
    \label{eq:rightprob}
\end{equation}
When $s_k$ is the minimum, the right side is maximally constrained: any replacement would need to be larger than $s_k$ but smaller than $s_{k+1}$, and such values are scarce.

The left constraint involves only the set $\{s_1, \ldots, s_{k-1}\}$. The right constraint involves only the disjoint set $\{s_{k+1}, \ldots, s_{\ell_{\max}}\}$. Because the quenched sequence is i.i.d., these two sets are \emph{statistically independent}.

A slot is rigid when \emph{at least one} of the two constraints is tight. For independent rare events, the probability of the union is approximately the sum of the individual probabilities:
\begin{equation}
    w_k = \mathbb{P}[\text{slot } k \text{ is rigid}] \approx \frac{1}{k} + \frac{1}{\ell_{\max}-k+1}\;.
    \label{eq:weight}
\end{equation}

The total effective number of two-level systems is the sum of the rigid-slot probabilities over all $\ell_{\max}$ positions:
\begin{equation}
    K(N) = \sum_{k=1}^{\ell_{\max}} w_k = \sum_{k=1}^{\ell_{\max}} \left(\frac{1}{k} + \frac{1}{\ell_{\max}-k+1}\right)\;.
    \label{eq:sum}
\end{equation}
The second sum is identical to the first (reindex $j = \ell_{\max}-k+1$), hence
\begin{equation}
    K(N) = 2 \sum_{k=1}^{\ell_{\max}} \frac{1}{k} = 2 H_{\ell_{\max}}\;,
    \label{eq:harmonic}
\end{equation}
where $H_n$ is the $n$-th harmonic number. Then, using $\ell_{\max} \sim 2\sqrt{N}$ and $H_n = \ln n + \gamma + O(1/n)$,
\begin{equation}
    K(N) = 2\bigl[\ln(2\sqrt{N}) + \gamma + o(1)\bigr] = \ln N + 2\ln 2 + 2\gamma + o(1)\;.
    \label{eq:asymptotic}
\end{equation}
Thus the effective number of two-level systems grows logarithmically: $K(N) \sim \ln N$.

\begin{acknowledgments}
We thank A. Hartmann for pointing to us the LIS problem and for inspiring discussions. R.M. thanks the LPTMS, Universit\'e Paris-Saclay where this work was in part accomplished. R. Mulet also acknowledges the support from the Marie Sklowdowskwa-Curie Actions (MSCA) Staff Exchange Project SIMBAD (REA grant agreement no. 101131463).
\end{acknowledgments}

\bibliographystyle{apsrev4-2}
\bibliography{referencesUlam}

@incollection{Ulam1961,
  author    = {S. M. Ulam},
  title     = {Monte Carlo calculations in problems of mathematical physics},
  booktitle = {Modern Mathematics for the Engineers},
  pages     = {261--281},
  year      = {1961},
  publisher = {McGraw-Hill}
}

@article{Bonomi2016,
  author  = {L. Bonomi and L. Xiong},
  title   = {On differentially private longest increasing subsequence computation in data stream},
  journal = {Transactions on Data Privacy},
  volume  = {9},
  number  = {1},
  pages   = {73--100},
  year    = {2016}
}

@article{Zhang2003,
  author  = {H. Zhang},
  title   = {Alignment of {BLAST} high-scoring segment pairs based on the longest increasing subsequence algorithm},
  journal = {Bioinformatics},
  volume  = {19},
  number  = {11},
  pages   = {1391--1396},
  year    = {2003}
}

@article{Prahofer2000,
  author  = {M. Pr{\"a}hofer and H. Spohn},
  title   = {Universal distributions for growth processes in $1+1$ dimensions and random matrices},
  journal = {Physical Review Letters},
  volume  = {84},
  number  = {21},
  pages   = {4882},
  year    = {2000}
}

@article{Baik1999,
  author  = {J. Baik and P. Deift and K. Johansson},
  title   = {On the distribution of the length of the longest increasing subsequence of random permutations},
  journal = {Journal of the American Mathematical Society},
  volume  = {12},
  number  = {4},
  pages   = {1119--1178},
  year    = {1999}
}

@article{Tracy1994,
  author  = {C. A. Tracy and H. Widom},
  title   = {Level-spacing distributions and the {A}iry kernel},
  journal = {Communications in Mathematical Physics},
  volume  = {159},
  number  = {1},
  pages   = {151--174},
  year    = {1994}
}

@article{Kardar1986,
  author  = {M. Kardar and G. Parisi and Y.-C. Zhang},
  title   = {Dynamic scaling of growing interfaces},
  journal = {Physical Review Letters},
  volume  = {56},
  number  = {9},
  pages   = {889},
  year    = {1986}
}

@article{Corwin2016,
  author  = {I. Corwin},
  title   = {{Kardar--Parisi--Zhang} universality},
  journal = {Notices of the American Mathematical Society},
  volume  = {63},
  number  = {3},
  pages   = {230--239},
  year    = {2016}
}

@book{MPV,
  author    = {M. M{\'e}zard and G. Parisi and M. A. Virasoro},
  title     = {Spin Glass Theory and Beyond},
  publisher = {World Scientific},
  year      = {1987}
}

@article{Krabbe2020,
  author  = {P. Krabbe and H. Schawe and A. K. Hartmann},
  title   = {Number of longest increasing subsequences},
  journal = {Physical Review E},
  volume  = {101},
  number  = {6},
  pages   = {062109},
  year    = {2020}
}

@article{Derrida1990,
  author  = {B. Derrida and O. Golinelli},
  title   = {Thermal properties of directed polymers in a random medium},
  journal = {Physical Review A},
  volume  = {41},
  number  = {8},
  pages   = {4160--4165},
  year    = {1990}
}

@article{Krabbe2023,
  author  = {P. Krabbe and H. Schawe and A. K. Hartmann},
  title   = {Replica symmetry breaking for {U}lam's problem},
  journal = {Physical Review B},
  volume  = {107},
  number  = {6},
  pages   = {064208},
  year    = {2023}
}

@article{Borjes2019,
  author  = {J. Borjes and H. Schawe and A. K. Hartmann},
  title   = {Large deviations of the length of the longest increasing subsequence of random permutations and random walks},
  journal = {Physical Review E},
  volume  = {99},
  number  = {4},
  pages   = {042104},
  year    = {2019}
}

\end{document}